\documentclass[sigconf]{acmart}

\usepackage{booktabs} 
\usepackage{amssymb}
\usepackage{amsmath}
\usepackage{amscd,amsfonts,amsbsy,rotating}
\usepackage{balance}
\usepackage{graphicx}
\usepackage{epsfig,epstopdf}
\usepackage{subfigure}
\usepackage{multirow}
\usepackage{booktabs}
\usepackage{color,xcolor}
\usepackage{url}
\usepackage{latexsym,bm}
\usepackage{enumitem,balance,mathtools}
\usepackage{wrapfig}
\usepackage{euscript}
\usepackage{algorithm}
\usepackage{algorithmic}
\usepackage{ifpdf}
\usepackage{diagbox}
\usepackage{caption}
\usepackage{makecell}
\usepackage{subfigure}
\usepackage{bm}

\newcommand{\bs}{\boldsymbol}

\newcommand{\btheta}{\bs{\theta}}

\AtBeginDocument{%
  \providecommand\BibTeX{{%
    \normalfont B\kern-0.5em{\scshape i\kern-0.25em b}\kern-0.8em\TeX}}}

\copyrightyear{2020} 
\acmYear{2020} 
\setcopyright{acmcopyright}\acmConference[SIGIR '20]{Proceedings of the 43rd International ACM SIGIR Conference on Research and Development in Information Retrieval}{July 25--30, 2020}{Virtual Event, China}
\acmBooktitle{Proceedings of the 43rd International ACM SIGIR Conference on Research and Development in Information Retrieval (SIGIR '20), July 25--30, 2020, Virtual Event, China}
\acmPrice{15.00}
\acmDOI{10.1145/3397271.3401440}
\acmISBN{978-1-4503-8016-4/20/07}

\begin{document}

\title{User Behavior Retrieval for Click-Through Rate Prediction}

\author{Jiarui Qin, Weinan Zhang, Xin Wu, Jiarui Jin, Yuchen Fang, Yong Yu}
\affiliation{
  \institution{Shanghai Jiao Tong University\\
    \{qinjr, wnzhang, wuxin, jerry, arthur\_fyc, yyu\}@apex.sjtu.edu.cn}
}

\fancyhead{}
\renewcommand{\shortauthors}{J. Qin, et al.}

\begin{abstract}
  Click-through rate (CTR) prediction plays a key role in modern online personalization services.
  In practice, it is necessary to capture user's drifting interests by modeling sequential user behaviors to build an accurate CTR prediction model. 
  However, as the users accumulate more and more behavioral data on the platforms, it becomes non-trivial for the sequential models to make use of the whole behavior history of each user. First, directly feeding the long behavior sequence will make online inference time and system load infeasible. Second, there is much noise in such long histories to fail the sequential model learning.
  The current industrial solutions mainly truncate the sequences and just feed recent behaviors to the prediction model, which leads to a problem that sequential patterns such as periodicity or long-term dependency are not embedded in the recent several behaviors but in far back history.
  To tackle these issues, in this paper we consider it from the data perspective instead of just designing more sophisticated yet complicated models and propose \textbf{U}ser \textbf{B}ehavior \textbf{R}etrieval for \textbf{CTR} prediction (UBR4CTR) framework. In UBR4CTR, the most relevant and appropriate user behaviors will be firstly retrieved from the entire user history sequence using a \emph{learnable} search method. These retrieved behaviors are then fed into a deep model to make the final prediction instead of simply using the most recent ones. It is highly feasible to deploy UBR4CTR into industrial model pipeline with low cost. Experiments on three real-world large-scale datasets demonstrate the superiority and efficacy of our proposed framework and models.
\end{abstract}



\maketitle

\section{Introduction} \label{sec:intro}
Click-through rate (CTR) prediction plays a key role in today's online personalization platforms (e.g., e-commerce, online advertising, recommender systems), whose goal is to predict a user's clicking probability on a specific item under a particular context. With more than a decade of development for the online personalization platforms, the amount of user behaviors logged on the platforms grows rapidly. There are 23\% of users have more than 1000 behaviors during six months on Taobao \cite{ren2019lifelong}. As there exist resourceful temporal patterns embedded in user behaviors, it becomes an essential problem for both industry and academia to build an effective and efficient model which could utilize user sequential behaviors to obtain accurate predictions of CTR.

In the deep learning era, there are many deep neural network (DNN) based CTR prediction models such as Wide\&Deep \cite{cheng2016wide}, FNN \cite{zhang2016deep}, DeepCross \cite{wang2017deep}, DeepFM \cite{guo2017deepfm}, PNN \cite{qu2016product,qu2018product} and xDeepFM \cite{lian2018xdeepfm}, most of which have been deployed on commercial personalization platforms. These models emphasize mining feature interactions, and are proposed to utilize the multi-categorical features of the data better. Nonetheless, such models ignore the sequential or temporal patterns of user behaviors.

As shown in \cite{hidasi2017recurrent,koren2009collaborative,he2016vista,agarwal2009spatio}, the temporal dynamics of user behavior play a key role in predicting the user's future interests. 
These sequential patterns include concept drifting \cite{widmer1996learning}, long-term behavior dependency \cite{koren2009collaborative,ren2019lifelong}, periodic patterns \cite{ren2018repeatnet}, etc.
Thus, there are models proposed to capture the user's sequential patterns in both CTR prediction and sequential recommendation tasks. For CTR prediction, there are attention-based models like DIN \cite{zhou2018deep} and DIEN \cite{zhou2019deep}, memory network-based models like HPMN \cite{ren2019lifelong}. More user behavior modeling methods are proposed for sequential recommendation, which is a quite similar task to CTR prediction. There are RNN-based models \cite{hidasi2015session}, CNN-based models \cite{tang2018personalized}, Transformer-based models \cite{kang2018self} and memory network-based models \cite{wang2018neural,Ebesu:2018:CMN:3209978.3209991}.

\begin{figure}[h]
    \centering
    \includegraphics[width=1\columnwidth]{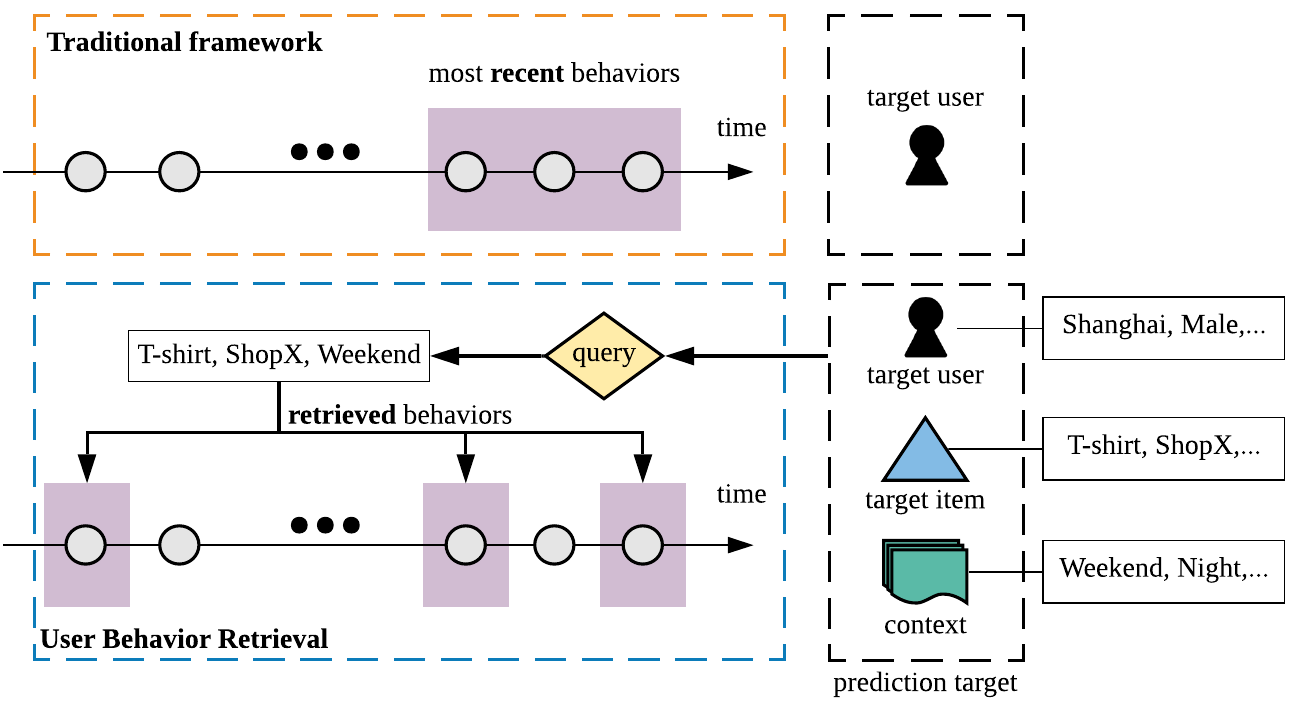}
    \caption{Comparison between traditional framework and UBR4CTR.}
    \label{fig:intro}
\end{figure}

However, most of the sequential models above have a common problem in real-world practice. When the platforms have logged a large number of user behaviors, the common industrial solution is to truncate the whole behavior sequence and only uses the most recent $N$ behaviors as the input to the prediction model \cite{zhou2018deep,zhou2019deep}, as illustrated in the upper part of Figure~\ref{fig:intro}. The strict requirements of online serving time plus bottleneck of system load and computational capacity put limits on the length of user sequence that could be used. As a result, in most of industrial cases, no more than 50 recent behaviors are used \cite{zhou2019deep}.

Traditional framework that uses the most recent $N$ behaviors could cause negative issues. It is obvious that the effective sequential patterns may be not just embedded in the recent sequence of behaviors. It may be traced back into further history such as periodicity and long-term dependencies \cite{ren2019lifelong}. If we try to use a longer sequence, however, a lot of irrelevant behaviors and noises could be introduced. Let alone the time and space complexity the longer history brings.

In this paper, to tackle the above practical issues, we try to solve the problem from the data perspective instead of designing more sophisticated yet complicated model. 
Specifically, we target to design a framework to retrieve a limited number of historic behaviors that are most useful for each CTR \textit{prediction target}. As shown in Figure~\ref{fig:intro}, a prediction target consists of three parts, i.e.,the target user, the target item and the corresponding context. There are features of the prediction target, such as the user's location, gender, occupation, and item's category, brand, merchant, and context features such as time and scenario. Then we use a model to select a subset of these features, which builds a query to retrieve the relevant historical behaviors. All the user's behaviors are stored as the information items in a search engine, and we use the generated query to search from the historical records. The retrieved behaviors are used in CTR prediction.

For every different candidate target item for the same user, we will retrieve different behaviors for prediction because the generated queries are different. This is a significant change compared with traditional framework that uses exactly the same recent $N$ behaviors for prediction on different items with the same user.

The resulted solution framework is called \textbf{U}ser \textbf{B}ehavior \textbf{R}etrieval for \textbf{CTR} (UBR4CTR). In UBR4CTR, the task is divided into two modules. The first is a \emph{learnable} retrieval module which consists of a self-attentive network to select the features and form the query, and a search engine in which the user behaviors are stored in an inverted index manner. The other module is the prediction module in which an attention-based deep neural network is built to make the final prediction based on the retrieved user behaviors as well as the features of the prediction target.

The contributions of the paper can be summarized in three-fold:
\begin{itemize}[leftmargin=15pt]
    \item We reveal an important fact that it is important to retrieve more relevant user behaviors than just use the most recent behaviors in user response prediction. Instead of designing more sophisticated and complex models, we put more attention on retrieving user's behavioral data.
    \item We propose a new framework called UBR4CTR which manages to retrieve different behaviors for the same user when predicting her CTR to different items under different contexts. All the previous sequential models only use the most recent behaviors of a user. We propose a search engine based method and an effective training algorithm to learn to retrieve appropriate behavioral data.
    \item We conduct extensive experiments and compare our framework with several strong baselines using traditional framework over three real-world large-scale e-commerce datasets. The results verify the efficacy of UBR4CTR framework.
\end{itemize}

The rest of the paper is organized as follows. In Section~\ref{sec:preli}, we will introduce the preliminaries and some notations used in this paper. Section~\ref{sec:method} is about the detailed description of our proposed framework and models. The experimental settings and corresponding results are shown in Section~\ref{sec:exp}. The deployment feasibility is discussed in Section~\ref{sec:deploy}. In Section~\ref{sec:related} we discuss about some important related works. Finally, we conclude the paper and discuss the future work in Section~\ref{sec:conclusion}.

\section{Preliminaries} \label{sec:preli}
In this section, we formulate the problem and introduce the notations. For CTR prediction task, there are $M$ users in $\mathcal{U} = \{u_1,..., u_M\}$ and $N$ items in $\mathcal{V} = \{v_1, ..., v_N\}$. The user-item interactions are denoted as $\mathcal{Y}=\{y_{uv} | u \in \mathcal{U}, v \in \mathcal{V} \}$ and,
\begin{equation}
	y_{uv} = \left\{
		\begin{array}{rcl}
			1, & & u~ \text{has clicked} ~v; \\
			0, & & \text{otherwise.} \\
		\end{array}
	\right.
\end{equation}
Furthermore, each user-item interaction has a timestamp and context in which the interaction happened, thus the data is formulated as quadruples which is $\{u, v, c, ts\}$ indicating $u$ clicked $v$ at time $ts$ in context $c$. 

To model user's evolving interests, we organize the user behavioral history as $H_u = \{b^u_1, b^u_2, ..., b^u_T\}$ in which $b^u_i$ stands for $i$-th behavior record of user $u$ sorting by timestamp. 

As click-through rate is essentially a matching probability between user, item and context, each behavior record $b^u_i$ is consist of these three parts that $b^u_i = [u, v_i, c_i]$ where $v_i$ is the $i$-th clicked item and $c_i$ is the context at which the interaction happened.

At the feature level, each user $u$ is represented by a bunch of features that $u = [f^u_1, ..., f^u_{K_u}]$ where $f^u_p$ denotes the $p$-th feature of $u$. It is a common practice that all the features are multiple categorical features \cite{zhou2018deep,ren2018bid}. Numerical features, if any, are discretized to categorical features. Similarly, $v = [f^v_1, ..., f^v_{K_v}]$ and $c = [f^c_1, ..., f^c_{K_c}]$. 
The goal of CTR prediction is to predict probability of target user $u$ clicking target item $v$ given historical behaviors of $u$ under context $c$. It is formulated as
\begin{equation}\label{eq:pred-func}
    \hat{y}_{uv} = \mathit{f}(u, v, c| H_u; \btheta),
\end{equation}
where $f$ is the learned function with parameters{} $\btheta$. We conclude the notations and the corresponding descriptions in Table~\ref{tab:notation}.

\begin{table}[t]
    \centering
    \caption{Notations and corresponding descriptions}\label{tab:notation}
    \resizebox{\columnwidth}{!}{
        \begin{tabular}{c|l}
            \hline
            Notation & Description. \\
            \hline
            $u, v$ & The target user and the target item. \\
            $M, N$ & The number of users and items. \\
            $y, \hat{y}$ & The label and the predicted probability of the user-item interaction. \\
            $H_u$ & The behavior sequence of $u$. \\
            $f^u_p, f^v_q, f^c_r$ & One feature of user, item and context. \\
            $\bm{u}, \bm{v}, \bm{c}$ & Embedding representation of target user $u$, target item $v$ and\\
            & the corresponding context $c$.\\
            $\bm{f}^u_p, \bm{f}^v_q, \bm{f}^c_r$ & Embedding representation of a feature. \\
            $\bm{b}^u_i$ & Embedding of $i$-th behavior of $u$.\\
            $B^u$ & The set of $u$'s retrieved behaviors. \\
            $S$ & Number of retrieved behaviors.\\
            \hline
        \end{tabular}
    }
    \vspace{-10pt}
\end{table}

\begin{figure*}[t]
	\centering
	\includegraphics[width=1.0\textwidth]{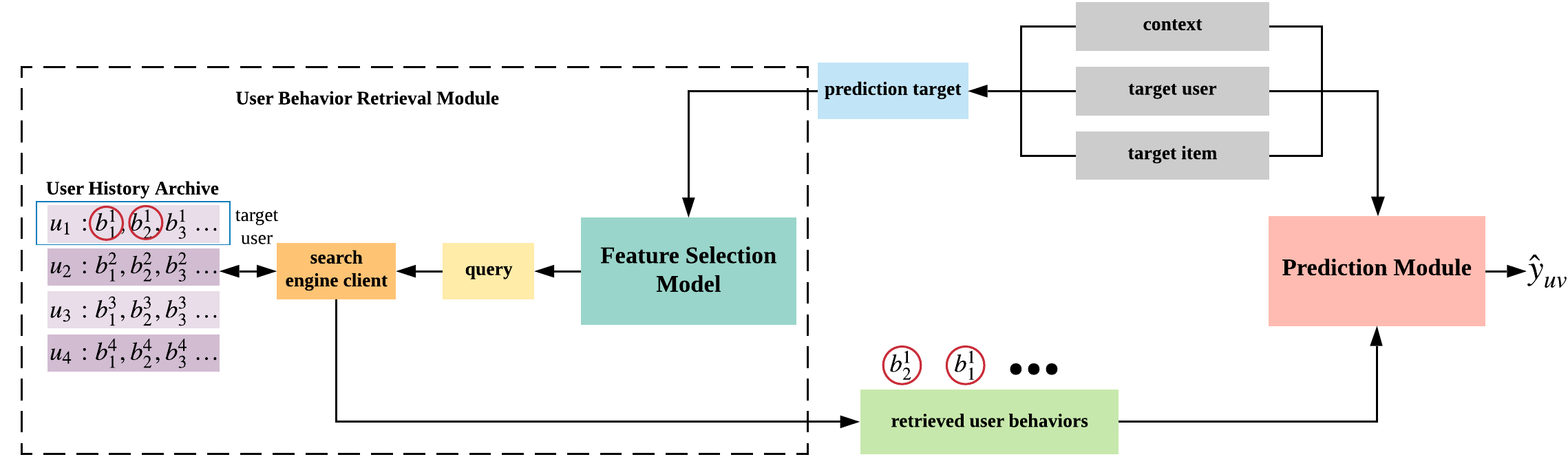}
	\caption{Overall framework of the proposed UBR4CTR framework.}
	\label{fig:framework}
\end{figure*}

\section{Methodology} \label{sec:method}
In this section, we describe our proposed UBR4CTR (\textbf{U}ser \textbf{B}ehavior \textbf{R}etrieval for \textbf{CTR} prediction) framework in detail. We firstly give a big picture of the overall framework, then we give detailed descriptions on the user behavior retrieval module and the prediction module. Moreover, the training methods and some analysis about time complexity are given in the following sections.

\subsection{Overall Framework}
The overall framework of UBR4CTR is illustrated in Figure~\ref{fig:framework}. The framework can be divided into two major modules: user behavior retrieval module and prediction module.

The user behavior retrieval module consists of a feature selection model, a search engine client and a user history archive. All the user historical behaviors are stored in the archive and they are organized in a feature-based inverted index manner which will be explained in details in Section~\ref{sec:search}. As shown in Figure~\ref{fig:framework}, when we need to predict the click-through rate between the target user and target item in a certain context, all of the three parts of information are combined to form a prediction target. The prediction target essentially consists of a bunch of features of the target user, target item and the context. So the prediction target is then fed to the feature selection model which will select the appropriate features to form a query. Detailed design of the feature selection model is in Section~\ref{sec:feat_select}. Then we use the query to search in the user history archive through a search engine client.

The search engine client retrieved a certain number of user behavior records and these records are then used by the prediction module. In the prediction module, we use an attention-based deep model to distinguish the influence of each behavior to the clicking probability and make the final prediction which will be discussed in Section~\ref{sec:prediction}.

The feature selection model and the prediction model are trained in turn. The goal of feature selection model is to select the most \textit{useful} subset of features. The features of this subset will be combined to generate a query which is used to retrieve the most relevant user behaviors for the final prediction.

\subsection{User Behavior Retrieval Module}
In this section, we introduce the user behavior retrieval module which consists of a feature selection model and a behavior searching process.

\subsubsection{Feature Selection Model} \label{sec:feat_select}
\begin{figure}[h]
    \centering
    \includegraphics[width=0.8\columnwidth]{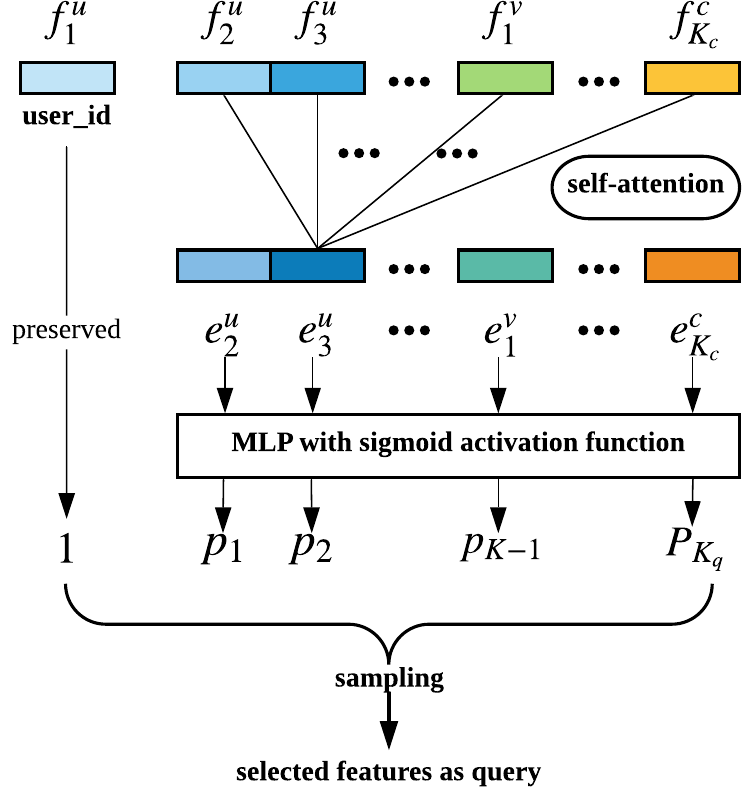}
    \caption{Illustration of feature selection model.}
    \label{fig:feature-selection-model}
\end{figure}

As shown in Figure~\ref{fig:feature-selection-model}, we regard all the features of target user $u$, target item $v$ and the corresponding context $c$ as the input of the feature selection model. Without loss of generality, we set $f^u_1$ as the user id feature. User id is a special feature which we must select because we want to retrieve behaviors of user $u$ herself. All the other features are concatenated as a whole. For simplicity, we denote all the features $[f^u_2, ..., f^u_{K_u}, f^v_1, ..., f^v_{K_v}, f^c_1, ..., f^c_{K_c}]$ as $[f_1, ..., f_{K_q}]$ correspondingly where $K_q=K_u + K_v + K_c - 1$.

To better model the relationships and interactive patterns between the features, we use self-attention mechanism \cite{vaswani2017attention}. Specifically, we define that
\begin{equation}
    Q = K = V = \left(
                  \begin{array}{c}
                          \bm{f}_{1} \\
                          \vdots \\
                          \bm{f}_{K_q}
                 \end{array}
                 \right)
\end{equation}
where $\bm{f}_i$ is the dense embedding representation of the $i$-th feature. And $K=Q=V \in R^{K_q \times d}$ where $d$ is the dimension of the embeddings.
The self-attention is defined as,
\begin{equation}
    \text{SA}(Q, K, V) = \operatorname{softmax}\left(\frac{Q K^{T}}{\sqrt{d_{}}}\right) V
\end{equation}
and multihead self-attention is
\begin{equation}
    E = \text{Multihead}(Q, K, V) = \text{Concat}(head_1, ..., head_h)W^O,
\end{equation}
where $head_i = \text{SA}(QW_i^Q, KW_i^K, VW_i^V)$. The parameters matrixes are $W^O \in R^{hd \times d}$, $W_i^Q \in R^{d \times d}$, $W_i^K \in R^{d \times d}$, $W_i^V \in R^{d \times d}$.

The output of multihead self-attention $E$ is then fed to a multi-layer perceptron with $ReLU(x)=max(0,x)$ activation function as
\begin{equation}
    \hat{E} = \text{MLP}(E),
\end{equation}
where $\hat{E} \in R^{K_q \times 1}$.
And the probability of selecting each corresponding feature is obtained by taking a sigmoid function as
\begin{equation}\label{eq:feat-prob}
    P = \left( 
          \begin{array}{c}
                  p_{1} \\
                  \vdots \\
                  p_{K_q}
         \end{array}
         \right) = \sigma(\hat{E}),
\end{equation}
where $\sigma(x)=\frac{1}{1+\exp^{-x}}$.

Then we sample the features according to these probabilities and thus get the selected subset of features. 
Note that the user id feature is always in the subset because we have to retrieve from the target user's own behaviors.

\subsubsection{Behavior Searching} \label{sec:search}
\begin{figure}[h]
    \centering
    \includegraphics[width=0.6\columnwidth]{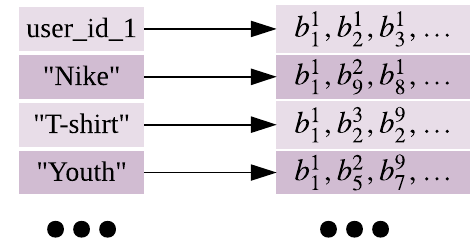}
    \caption{Illustration of feature based inverted index.}
    \label{fig:feature-based-inverted-index}
\end{figure}

We use the typical search engine approach to store and retrieve the user behaviors. We regard each user behavior as a document and each feature as a term. Inverted index is used to track all the behaviors, which is shown in Figure~\ref{fig:feature-based-inverted-index}. Each feature value is associated with a posting list which consists of all the behaviors that has this specific feature value. For instance, the "user\_id\_1" has the posting list of all the behaviors of the user whose id is 1 and "Nike" has the posting list which consists of the behavior records that brand ids are "Nike".

The logic of the query is essentially formulated as
\begin{equation}\label{eq:query-form}
    f_1^u~AND~(f_1~OR~f_2~OR ...OR~f_n),    
\end{equation}
where $f_1^u$ is user id. The selected query feature set is $q = \{f_1, f_2, ...f_n\}$. The posting list of $f_1^u$ and the union set of $f_1$ to $f_n$'s posting lists are intersected. The intersection is the candidate behavior set. Then we use BM25 \cite{robertson1995okapi} to score every behavior document in the candidate set and the top $S$ are retrieved.


The similarity $s$ between the query $q$ and a behavior document $D$ is calculated as,
\begin{equation}
    s = \sum_{i=1}^{n} \operatorname{IDF}\left(f_{i}\right) \cdot \frac{tf\left(f_{i}, D\right) \cdot\left(k_{1}+1\right)}{tf\left(f_{i}, D\right)+k_{1} \cdot\left(1-b+b \cdot \frac{|D|}{\text { avgdl }}\right)}, 
\end{equation}
where $tf\left(f_{i}, D\right)$ is feature $f_i$'s term frequency in $D$ which is 1 or 0.
The length of a behavior document is defined as the number of the features in that behavior, so all the behavior documents have the same length. Thus the average length is each documents' length and $\frac{|D|}{\text{avgdl}}=1$. $k_1$ and $b$ are free parameters. We set $k_1=1.2$ and $b=0.75$.
IDF is defined as,
\begin{equation}
    \operatorname{IDF}\left(f_{i}\right)=\log \frac{\mathcal{N}-\mathcal{N}\left(f_{i}\right)+0.5}{\mathcal{N}\left(f_{i}\right)+0.5},
\end{equation}
where $\mathcal{N}$ is the total number of the behavior documents and $\mathcal{N}\left(f_{i}\right)$ is the number of documents that contains the feature $f_i$. The IDF term gives the common features less importance than rare features. It makes sense that rare features imply stronger signals on user's preference compared to the commonly seen features.

By ranking the candidate behaviors using BM25, we obtain the top $S$ documents as the retrieved behaviors.

\subsection{Prediction Module}\label{sec:prediction}
For the prediction module, we use an attention-based deep neural network to model the importance of different user behaviors to the final prediction.

\begin{figure}[h]
    \centering
    \includegraphics[width=0.9\columnwidth]{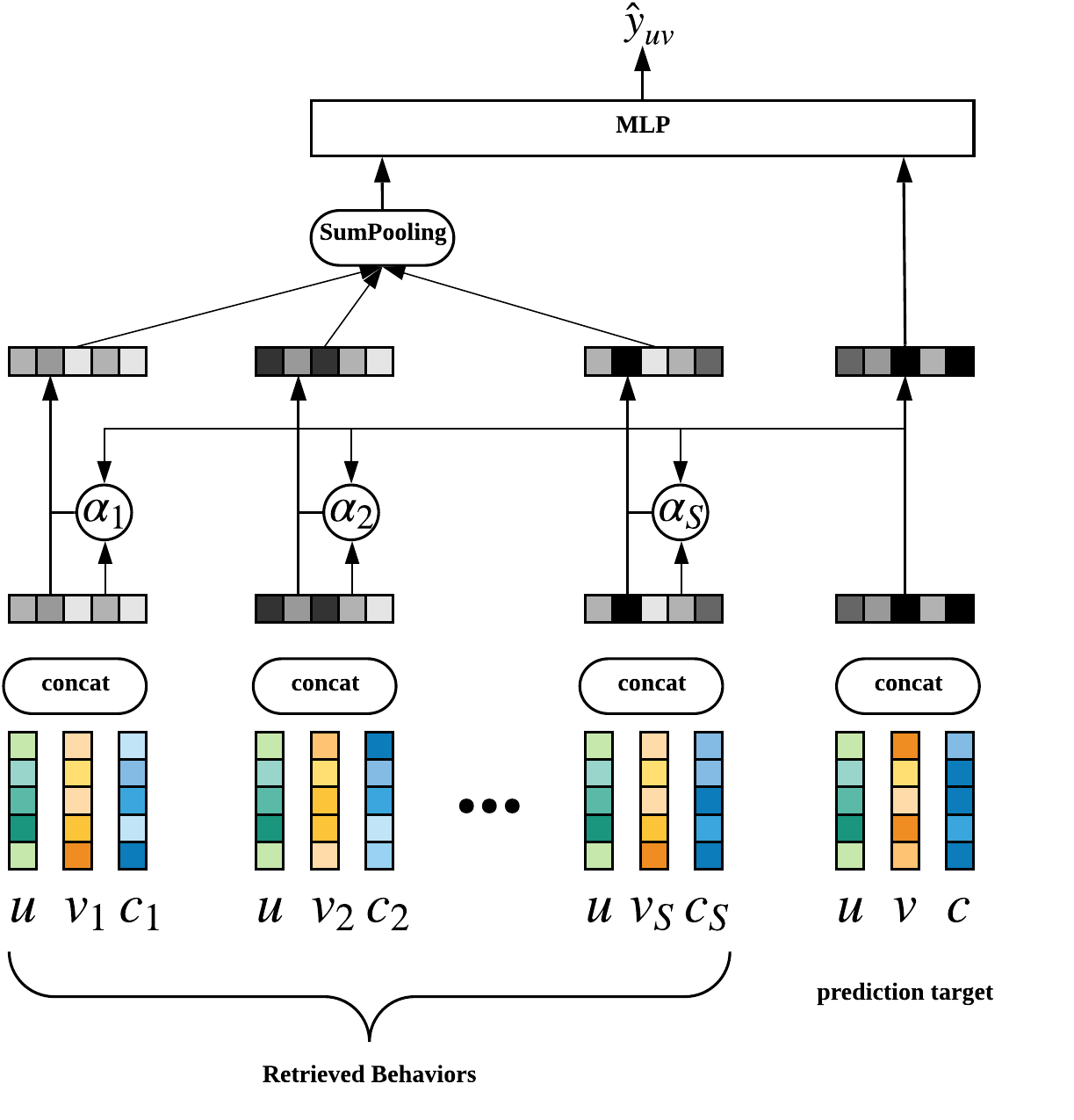}
    \caption{Illustration of attention-based prediction network of the prediction module.}
    \label{fig:atten-pred-net}
\end{figure}

As shown in Figure~\ref{fig:atten-pred-net}, the comprehensive user representation $\bm{r^u}$ is calculated by weighted sum pooling as
\begin{equation} \label{eq:weight-sum}
    \bm{r}^u = \sum_{i=1}^S \alpha_i \cdot \bm{b}^u_i, \bm{b}^u_i \in B^u
\end{equation}
where $\bm{b}^u_i = [\bm{u}, \bm{v}_i, \bm{c}_i]$ and $\alpha_i$ represents the contribution of $b^u_i$ to the comprehensive user representation. The attention weight $\alpha_i$ is calculated as
\begin{equation} 
    \alpha_i = \frac{\exp(w_i)}{\sum_{j=1}^S \exp(w_j)},
\end{equation}
where $w_i$ is defined as,
\begin{equation}\label{eq:att}
    w_i = \text{Att}(\bm{b}^u_i, \bm{t}),
\end{equation}
and prediction target $\bm{t} = [\bm{u}, \bm{v}, \bm{c}]$.
$\text{Att}$ is a multi-layer deep network with ReLU activation function.

Then the final prediction is calculated as
\begin{equation}
    \hat{y}_{uv}= f_{\phi}\left(B^u, \bm{u}, \bm{v}, \bm{c}\right) = f_{\phi}\left(\bm{r^u}, \bm{u}, \bm{v}, \bm{c}\right)
\end{equation}
where $f$ is implemented as a three-layer perceptron, whose widths are 200, 80 and 1 respectively. $\phi$ is the parameters. The output layer uses sigmoid function and other layers use ReLU as activation.

\subsection{Model Training}
As the objective is to estimate the click-through rate accurately, we utilize the log-likelihood as our objective function which is defined as,
\begin{equation}\label{eq:obj}
    J^{\pi_{\theta}, f_{\phi}} = \max_{\theta} \max_{\phi} \sum_u \sum_v E_{B^u \sim \pi_{\theta}(B^u|q)} [LL(y_{uv}, f_{\phi}(B^u, u, v, c))],
\end{equation}
where $LL$ is the log-likelihood of predicted score on target user-item pair $u$, $i$ given retrieved user behaviors $B^u$ under context $c$. It is defined as,
\begin{equation} \label{eq:logloss}
    \begin{aligned}
    LL(y_{uv}, f_{\phi}(B^u, u, v, c)) & = y_{uv} \cdot \log(f_{\phi}(B^u, u, v, c)) \\
    & + (1-y_{uv}) \cdot \log(1-f_{\phi}(B^u, u, v, c)).
    \end{aligned}
\end{equation}
In Eq.~\ref{eq:obj}, $q$ is the query which consists of the selected features as described in Section~\ref{sec:feat_select}. Sampling is used to select features and then they form the query string. After the query string is formed, searching results, a.k.a, retrieved behaviors are deterministic. Thus we could regard the behaviors are sampled from a probability distribution $\pi_{\theta}(B^u|q)$.

\subsubsection{Optimize the Prediction Module}
To optimize the attention-based prediction network $f_{\phi}$, we could have that,
\begin{equation}\label{eq:pred-optimization}
  \begin{aligned}
    \phi^* &= \text{arg}\min_{\phi} (L_{ce} + \lambda L_r) \\
    & =  \text{arg}\min_{\phi} \sum_u \sum_v E_{B^u \sim \pi_{\theta}(B^u|q)} [-LL(y_{uv}, f_{\phi}(B^u, u, v, c))] \\
    & + \frac{1}{2} \lambda \left( \| \bm{\Phi} \|_2^2 \right) ~,
  \end{aligned}
\end{equation}
where $L_{ce}$ is cross entropy loss and $L_r$ is regularization term.
When we are optimizing the prediction network, the retrieval module remains unchanged, so if the function $f$ is differential with respect to $\phi$, the above optimization can be solved by typical stochastic gradient descent algorithm.

\subsubsection{Optimize the Retrieval Module}
For the retrieval module, only the feature selection model needs to be optimized. As a sampling process is evolved, we could not directly use SGD to optimize it. Instead, we use the REINFORCE \cite{williams1992simple, yu2017seqgan} algorithm to deal with the discrete optimization.

Specifically, while keeping the prediction network $f_{\phi}$ fixed, the feature selection model is optimized via performing its maximization:
\begin{equation}\label{eq:retrieve-obj}
    \theta^* = \text{arg}\max_{\theta} \sum_u \sum_v E_{B^u \sim \pi_{\theta}(B^u|q)} [LL(y_{uv}, f_{\phi}(B^u, u, v, c))],
\end{equation}
we denote $LL(y_i, f_{\phi}(B^K_i, u_i, i_i, c_i))$ as $LL(\cdot)$ because it doesn't have parameter $\theta$. For each query $q$, we denote objective function as $J^{q}$ which is
\begin{equation}\label{eq:retrieve-obj-q}
    J^{q} = E_{B^u \sim \pi_{\theta}(B^u|q)} [LL(\cdot)].
\end{equation}
We regard the retrieval module $\pi_{\theta}(B^u|q)$ as a policy and use the likelihood ratio to estimate its gradients as,
\begin{equation}\label{eq:retrieve-optimization}
    \begin{aligned}
    \nabla_{\theta}(J^{q}) &=  \nabla_{\theta} E_{B^u \sim \pi_{\theta}(B^u|q)} [LL(\cdot)] \\
    &= \sum_{B^u_i \in \mathcal{B}} \nabla_{\theta} \pi_{\theta}(B^u_i|q) [LL(\cdot)] \\
    &= \sum_{B^u_i \in \mathcal{B}} \pi_{\theta}(B^u_i|q) \nabla_{\theta} \log(\pi_{\theta}(B^u_i|q)) [LL(\cdot)]\\
    &= E_{B^u \sim \pi_{\theta}(B^u|q)} \nabla_{\theta} \log(\pi_{\theta}(B^u|q)) [LL(\cdot)]\\
    & \simeq \frac{1}{L} \sum_{l=1}^L \nabla_{\theta} \log(\pi_{\theta}(B^u_l|q)) [LL(\cdot)],
    \end{aligned}
\end{equation}
which is an unbiased estimation on the gradients of Eq.~\ref{eq:retrieve-obj-q}.

As the uncertainty of the entire retrieval module actually comes from the feature selection model, we could derive that,
\begin{equation}
    \pi_{\theta}(B^u|q) = \prod_{j=1}^n p(f_j),
\end{equation}
where $f_j \in \{f_1,..,f_n\}$ and $p(f_j)$ is the sampling probability obtained in Eq.~\ref{eq:feat-prob}.
So in Eq.~\ref{eq:retrieve-optimization} the gradient of $\theta$ can be further derive as
\begin{equation}
    \nabla_{\theta}(J^{q}) \simeq \frac{1}{L} \sum_{l=1}^L \sum_{j=1}^n \nabla_{\theta} \log(p(f^l_j)) [LL(\cdot)],
\end{equation}
where $f^l_j$ is $j$-th feature of $l$-th query. To train the model with a reward that has a better scale, we replace $LL(\cdot)$ with Relative Information Gain (RIG) as the reward function here. RIG is defined as $RIG = 1 - NE$ where NE is the normalized entropy \cite{he2014practical}.
NE is calculated as,
\begin{equation}
    NE = \frac{LL(\cdot)}{p \log (p)+(1-p) \log (1-p)}
\end{equation}
where $p$ is the average experienced CTR.

\subsubsection{Pseudo Code of Training Process.}
In this subsection, we give a detailed pseudo code of the training process. First we pre-train the prediction network with the initial feature selection model. After the pre-train, the two models are optimized in turn. Algorithm~\ref{algo:training} shows the training process.
\begin{algorithm}[!h]
        \caption{Training the UBR4CTR framework}
        \label{algo:training}
        \begin{algorithmic}[1]
            \REQUIRE
            Dataset $\mathcal{D} = (\mathcal{U}_{target}, \mathcal{V}_{target}, \mathcal{C}_{target})$ containing all the target user-item-context triples; User history archive $H_u$.
            \ENSURE
            final CTR prediction $\hat{Y}$ between all the target user $u$ and target item $v$.
            \vspace{1mm}
            \STATE Initialize all parameters.
            \STATE Select the features and form the queries $\mathcal{Q} = \{q,...\}$ for each prediction target $[u, v, c] \in \mathcal{D}$ using the initialized feature selection model.
            \STATE Obtain the retrieved behaviors $\mathcal{B}=\{B^u,...\}$ of the queries $\mathcal{Q}$ using the search engine as described in Section~\ref{sec:search}.
            \STATE Train the attention-based prediction network using Eq.~\ref{eq:pred-optimization} for one epoch.
            \REPEAT 
            \STATE Train retrieval model using Eq.~\ref{eq:retrieve-obj} for one epoch.
            \STATE Select the features and form the queries $\mathcal{Q} = \{q,...\}$ for each prediction target $[u, v, c] \in \mathcal{D}$ using the feature selection model.
            \STATE Obtain the retrieved behaviors $\mathcal{B}=\{B^u,...\}$ of the queries $\mathcal{Q}$ using the search engine as described in Section~\ref{sec:search}.
            \STATE Train attention-based prediction network using Eq.~\ref{eq:pred-optimization} for one epoch.
            \UNTIL convergence
        \end{algorithmic}
    \end{algorithm}

\subsection{Model Analysis}\label{sec:time-comp}
In this section, we analyze the time complexity and feasibility of our method.
We use $\mathcal{N}$ to denote the total number of all users' logged behaviors and use $F$ to denote the total number of unique features (equivalent to term) that have ever appeared in the whole dataset. Then the average length of the posting lists in the user history archive is $\frac{\mathcal{N}}{F}$. Recall the searching operation described in Section~\ref{sec:search} and Eq.~\ref{eq:query-form}, we first retrieve all the postings of features in $q$ which takes $O(1)$ time. Then the interaction operation takes $O(T + K_q \cdot \frac{\mathcal{N}}{F})$ time where $T$ is the average length of a user sequence and $K_q = K_u + K_v + K_c - 1$ is the upper bound of the number of selected features. The next scoring operations does not increase the complexity because it is linear to $O(T + K_q \cdot \frac{\mathcal{N}}{F})$. 
The complexity of the self-attention in feature selection model is $O(K_q^2)$. The attention-based prediction network takes $O(C)$ time where $C$ is the cost of computing one Att operation in Eq.~\ref{eq:att} because all the attention operations can be paralleled. These two constants can be ignored and the total time  complexity of UBR4CTR is $O(T + K_q \cdot \frac{\mathcal{N}}{F})$.


\section{Experiments} \label{sec:exp}
In this section, we present our experimental settings and corresponding results in detail. We compare our model with several strong baselines and achieve the state-of-the-art performance. Furthermore, we have published our code for reproduction\footnote{https://github.com/qinjr/UBR4CTR}.

We start with three research questions (RQ) and use them to lead the following discussions.
\begin{itemize}
  \item \textbf{RQ1} Does UBR4CTR achieves the best performance compared to other baselines?
  \item \textbf{RQ2} What is the convergence performance of Algorithm~\ref{algo:training}? Is the training process effective and stable?
  \item \textbf{RQ3} What is the influence of the retrieval module in UBR4CTR and how the retrieval size affect the performance?
\end{itemize}

\subsection{Experimental Settings}
\subsubsection{Datasets}
We use three real-world and large-scale datasets of users online behaviors from three different platforms of Alibaba Group. The statistics of the datasets can be found in Table~\ref{tab:stats}.

\begin{description}[leftmargin=15pt]
  \item [Tmall]\footnote{https://tianchi.aliyun.com/dataset/dataDetail?dataId=42} is provided by Alibaba Group which contains user behavior history on Tmall e-commerce platform from May 2015 to November 2015.
  \item [Taobao] \cite{zhu2018learning} is a dataset consisting of user behavior data retrieved from Taobao\footnote{https://tianchi.aliyun.com/dataset/dataDetail?dataId=649}, one of the biggest e-commerce platforms in China. It contains user behaviors from November 25 to December 3, 2017, with several behavior types including click, purchase, add to cart and item favoring.
  \item [Alipay]\footnote{https://tianchi.aliyun.com/dataset/dataDetail?dataId=53} is collected by Alipay which is an online payment application. The users online shopping behaviors are from July 1, 2015 to November 30, 2015.
\end{description}

\begin{table}[t]
  \centering
  \caption{The dataset statistics.}\label{tab:stats}
  \resizebox{\columnwidth}{!}{
    \begin{tabular}{c|c|c|c|c|c}
      \hline
      Dataset & Users \# & Items \# & Interaction \# & Avg. Seq. Length & Feature Field \#\\
      \hline
      Tmall & 424,170 & 1,090,390 & 54,925,331 & 129 & 9\\
      \hline
      Taobao & 987,994 & 4,162,024 & 100,150,807 & 101 & 4\\
      \hline
      Alipay & 498,308 & 2,200,291 & 35,179,371 & 70 & 6\\
      
      \hline
    \end{tabular}
  }
\end{table}

\textbf{Dataset Preprocessing.}
For UBR4CTR, the datasets are processed into the format of comma separated features. A line containing user, item and context features is treated as a behavior document. For baselines, the user behaviors are simply sorted by timestamp.
As the datasets don't contain specific context features, we manually design some features using behavior timestamp to make it possible to capture periodicity. We design features such as season id (spring, summer, etc), weekend or not, and which half of the month it is.

\textbf{Search Engine.}
After the datasets are preprocessed, they are inserted into a search engine using a comma separated tokenizer. We use Elastic Search \footnote{https://www.elastic.co} as the search engine client which is based on Apache Lucene\footnote{http://lucene.apache.org}.

\textbf{Train \& Test Splitting.}
We split the datasets using the time step. The training dataset contains the 1st to $(T-2)$th user behaviors, in which the 1st to $(T-3)$th behaviors are used to predict the behavior at $(T-2)$th step. Similarly, the validation set uses 1st to $(T-2)$th behaviors to predict $(T-1)$th behavior and the test set uses 1st to $(T-1)$th behaviors to predict $T$th behavior.

\textbf{Hyperparameters.}
The learning rate of feature selection model of UBR4CTR is searched from $\{1 \times 10^{-6}, 1 \times 10^{-5}, 1 \times 10^{-4}\}$, learning rate for attention based prediction network is from $\{1 \times 10^{-4}, 5 \times 10^{-4}, 1 \times 10^{-3}\}$ and the regularization term is from $\{1 \times 10^{-4}, 5 \times 10^{-4}, 1 \times 10^{-3}\}$. The search space of learning rate and regularization term for baseline models are the same with prediction network in UBR4CTR. Batch size is from $\{100, 200\}$ for all models. The hyperparameters of each model are tuned and the best performances have been reported in Section~\ref{sec:results}.

\subsubsection{Evaluation Metrics}
We evaluate the CTR prediction performance with two widely used metrics. The first one is area under ROC curve (AUC) which reflects the pairwise ranking performance between click and non-click samples. The other metric is log loss. Log loss is to measure the overall likelihood of the test data and has been widely used for the classification tasks \cite{ren2016user,ren2018bid}.

\subsubsection{Compared Baselines}
We compare our framework and models with seven different strong baselines from both sequential CTR prediction and recommendation scenarios.

\begin{itemize}[leftmargin=40pt]
  \item [\textbf{GRU4Rec}] \cite{hidasi2015session} is based on GRU and it is the first work using the recurrent cells to model sequential user behaviors for session-based recommendation.
  \item [\textbf{Caser}] \cite{tang2018personalized} is a CNNs-based model that regards the user sequence as an image thus uses horizontal and vertical convolutional layers to capture temporal patterns of user behaviors.
  \item [\textbf{SASRec}] \cite{kang2018self} uses Transformer \cite{vaswani2017attention}. It regards the user behaviors as a sequence of tokens and uses self-attention mechanism and position embedding to capture the dependencies and relations between behaviors.
  \item [\textbf{HPMN}] \cite{ren2019lifelong} is a hierarchical periodic memory network that is proposed to handle very long user historical sequence. Moreover, the user memory state can be updated incrementally.
  \item [\textbf{MIMN}] \cite{pi2019practice} is based on Neural Turing Machine \cite{graves2014neural} which models multiple channels of user interests drifting. The model is implemented as a part of user interest center \cite{pi2019practice} which could model very long user behavior sequences.
  \item [\textbf{DIN}] \cite{zhou2018deep} is the first model that uses attention mechanism in CTR prediction of online advertising.
  \item [\textbf{DIEN}] \cite{zhou2019deep} uses two-layer RNNs with attention mechanism to capture evolving user interests. It uses the calculated attention values to control the second recurrent layer, which is called AUGRU.
  \item [\textbf{UBR4CTR}] is our proposed framework and models described in Section~\ref{sec:method}.
\end{itemize}

\subsection{Performance Comparison: RQ1} \label{sec:results}

We conduct two groups of comparisons between our UBR4CTR and baseline models. In the first group of experiments, all of the models are using the same amount of user behaviors which are 20, 20, 12 for the three datasets respectively. The only difference is that baselines are using the most recent behaviors (about 20\% of the total length) and UBR4CTR retrieve 20\% of behaviors from the whole sequence. The experimental results are shown in Table~\ref{tab:ctr_short}.

From the table, we could find the following facts.
(i) The performance is improved significantly compared to the baselines. AUC are improved by 4.1\%, 10.9\% and 22.3\% on three datasets respectively, and log-loss are improved by 9.0\%, 12.0\% and 32.3\% respectively. 
(ii) The vast improvement is a demonstration that the most recent behaviors do not embed enough temporal patterns so the baselines can not capture them effectively. Although some of the baselines are pretty complex and fancy, they cannot perform well if the patterns that they try to capture are not contained in the recent behavior sequence in the first place.

\begin{table}
  \centering
  \caption{The first group of results of CTR prediction in terms of AUC (the higher, the better) and log-loss (LL, the lower, the better). Bold values are the best in each column, while the second best values are underlined. Improvements are against the second best results.}
  \label{tab:ctr_short}
  \begin{tabular}{c|cc|cc|cc}
    \hline
    \multirow{2}{*}{ Model } & \multicolumn{2}{c|}{Tmall} & \multicolumn{2}{c|}{Taobao} & \multicolumn{2}{c}{Alipay} \\
    & AUC & LL & AUC & LL & AUC & LL \\
    \hline
    GRU4Rec & 0.762 & 0.585 & 0.677 & 0.661 & 0.6131 & 0.699 \\
    \hline
    Caser & 0.762 & 0.579 & 0.673 & 0.657 & 0.655 & 0.676 \\
    \hline
    SASRec & 0.755 & 0.586 & 0.670 & 0.658 & 0.648 & 0.679 \\
    \hline
    HPMN & 0.763 & 0.579 & 0.668 & 0.660 & 0.615 & 0.703 \\
    \hline
    MIMN & 0.753 & 0.591 & 0.662 & 0.686 & 0.664 & 0.675 \\
    \hline
    DIN & 0.766 & 0.576 & \underline{0.678} & \underline{0.649} & \underline{0.732} & \underline{0.616} \\
    \hline
    DIEN & \underline{0.775} & \underline{0.567} & 0.677 & 0.659 & 0.730 & \underline{0.616} \\
    \hline
    UBR4CTR & \textbf{0.807} & \textbf{0.516} &\textbf{0.752} & \textbf{0.571} & \textbf{0.895} & \textbf{0.417} \\
    \hline
    \hline
    Imprv. & 4.1\% & 9.0\% & 10.9\% & 12.0\% & 22.3\% & 32.3\%
  \end{tabular}
\end{table}

\begin{table}
  \centering
  \caption{The second group of results of CTR prediction in terms of AUC (the higher, the better) and log-loss (LL, the lower, the better). Improvements are against the second best results.}
  \label{tab:ctr_long}
  \begin{tabular}{c|cc|cc|cc}
    \hline
    \multirow{2}{*}{ Model } & \multicolumn{2}{c|}{Tmall} & \multicolumn{2}{c|}{Taobao} & \multicolumn{2}{c}{Alipay} \\
    & AUC & LL & AUC & LL & AUC & LL \\
    \hline
    GRU4Rec & 0.781 & 0.560 & 0.677 & 0.660 & 0.639 & 0.684 \\
    \hline
    Caser & 0.774 & 0.566 & 0.645 & 0.659 & 0.705 & 0.631 \\
    \hline
    SASRec & 0.769 & 0.578 & 0.669 & \underline{0.654} & 0.711 & 0.637 \\
    \hline
    HPMN & 0.767 & 0.579 & 0.655 & 0.664 & 0.703 & 0.643 \\
    \hline
    MIMN & 0.759 & 0.590 & 0.659 & 0.659 & 0.719 & 0.634 \\
    \hline
    DIN & 0.791 & 0.546 & 0.605 & 0.679 & \underline{0.856} & 0.506 \\
    \hline
    DIEN & \underline{0.805} & \underline{0.538} & \underline{0.704} & 0.656 & 0.843 & \underline{0.491} \\
    \hline
    UBR4CTR & \textbf{0.807} & \textbf{0.516} &\textbf{0.752} & \textbf{0.571} & \textbf{0.895} & \textbf{0.417} \\
    \hline
    \hline
    Imprv. & 0.2\% & 4.1\% & 6.8\% & 12.7\% & 4.6\% & 15.1\%
  \end{tabular}
\end{table}

In the second group of experiment, we use different settings for the baselines and exactly the same settings for UBR4CTR as the first group of experiment. We feed the full-length sequences to all the baseline models which are 100, 100 and 60 respectively on three datasets. They are the maximum lengths of user behaviors in these datasets. And the sizes of the retrieved behaviors remain the same for UBR4CTR which are 20, 20 and 12. The results are shown in Table~\ref{tab:ctr_long}.

In Table~\ref{tab:ctr_long}, the performance of UBR4CTR is the same as that in Table~\ref{tab:ctr_short} because we don't change any settings.

From the table, we could find the following facts. 
(i) UBR4CTR still has the best performance even though it uses $80\%$ less behaviors than other baselines. This shows that longer sequence may has more noise and irrelevant information thus it is necessary to obtain only the most useful data out of the whole sequence.
(ii) Most of the baselines achieve better performance than themselves compared to Table~\ref{tab:ctr_short}, especially DIN and DIEN. This shows that behaviors from further history do contain richer information and patterns. And these patterns are easier to be captured using attention mechanism.
(iii) Although the improvement of AUC is much smaller due to the better performance of baselines, log-loss still improves significantly. The reason is that the optimization objective of the retrieval module is RIG (equivalent to log-loss) after all.

\subsection{Learning Process: RQ2}
\begin{figure}[h]
    \centering
    \includegraphics[width=1.0\columnwidth]{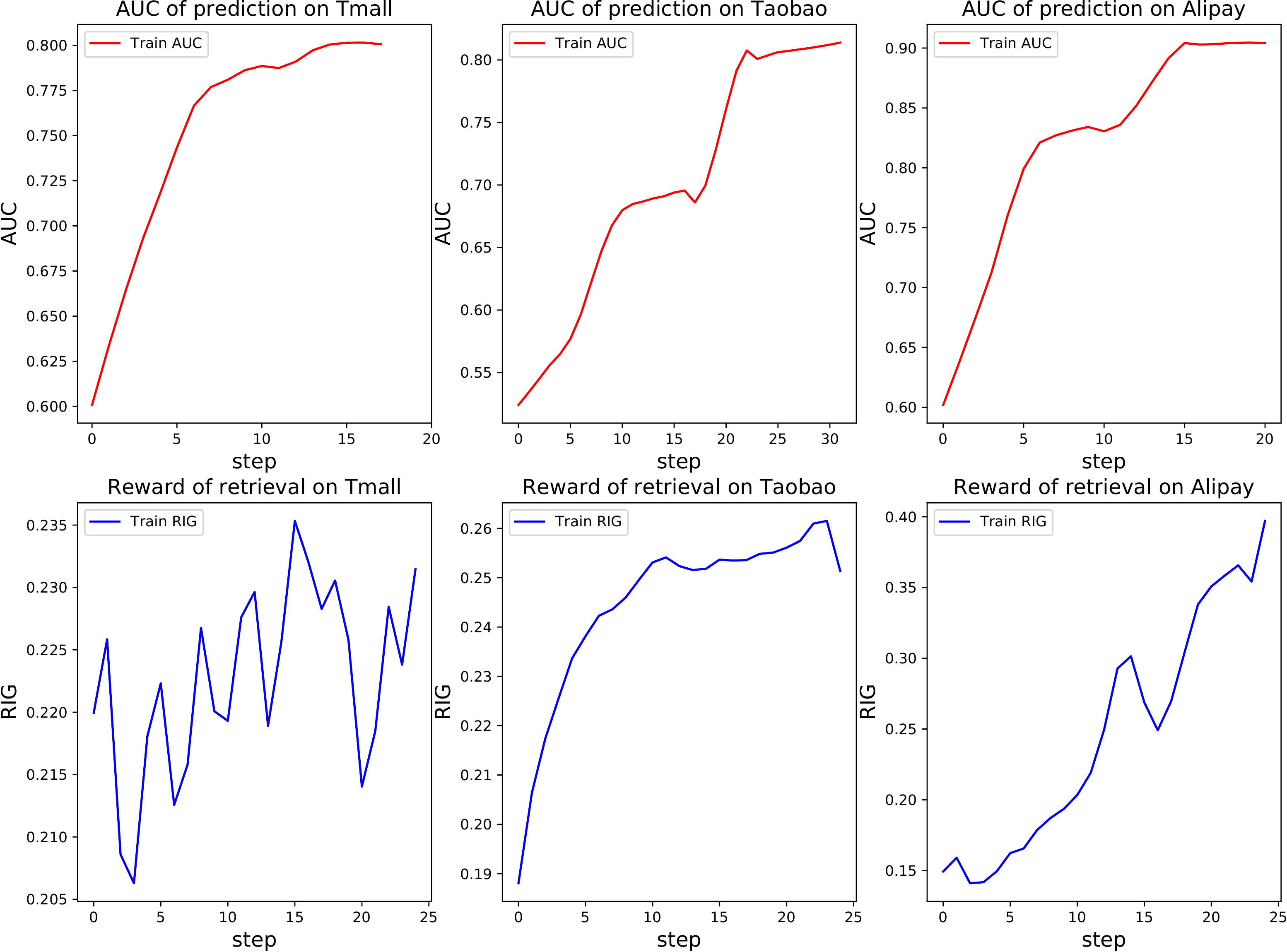}
    \caption{Learning curves of UBR4CTR.}
    \label{fig:learning_curve}
\end{figure}

To illustrate the convergence of our framework, we plot the learning curves of UBR4CTR. 
In Figure~\ref{fig:learning_curve}, the upper three subplots are the AUC curves of the attentive prediction network when training on three datasets, respectively. Each step of the x-axis is corresponding to the iteration over $4\%$ of the training set.

The lower three sub-figures show the "reward" of the REINFORCE algorithm w.r.t the feature selection model of the retrieval module. The "reward" is essentially RIG which is a variant of log-likelihood. Each step of the x-axis means the iteration over $4\%$ of the training set. The rewards increase during the training process implying that the feature selection model actually learns useful patterns.

From the AUC figures, we could find that our models converge effectively. For the prediction network, we can observe that there are flat areas in the middle of training process followed by a rapid increase, especially in the second and third AUC plots. Recall our training procedure described in Algorithm~\ref{algo:training}, the retrieval module is trained after the prediction network. It means that when the prediction network is about to converge, the retrieval module begins training, and after that, there will be a performance break though for the prediction network.

\subsection{Extensive Study: RQ3}
\begin{figure}[h]
    \centering
    \includegraphics[width=1.0\columnwidth]{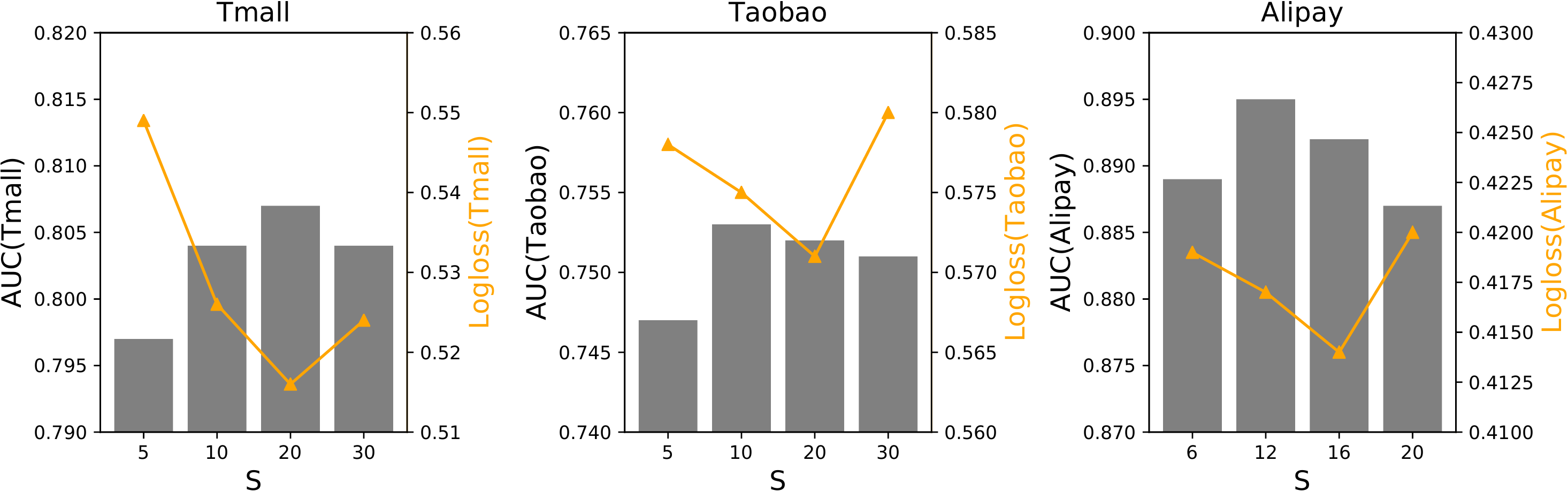}
    \caption{Influence of different retrieval size.}
    \label{fig:s_study}
\end{figure}

In this section, we conduct some extensive and ablation studies on our framework. 
Figure~\ref{fig:s_study} illustrates the influence of different retrieval sizes on the prediction performance. From the figure, we can find that the fluctuation of AUC and log-loss is not very severe in terms of the absolute values. However, there exists an optimal size for each dataset. This reveals that smaller sizes may not contain enough behaviors and information, while too much retrieved behaviors are not always suitable for performance either. Because it will introduce too much noise.

\begin{figure*}[t]
  \centering
  \includegraphics[width=0.9\textwidth]{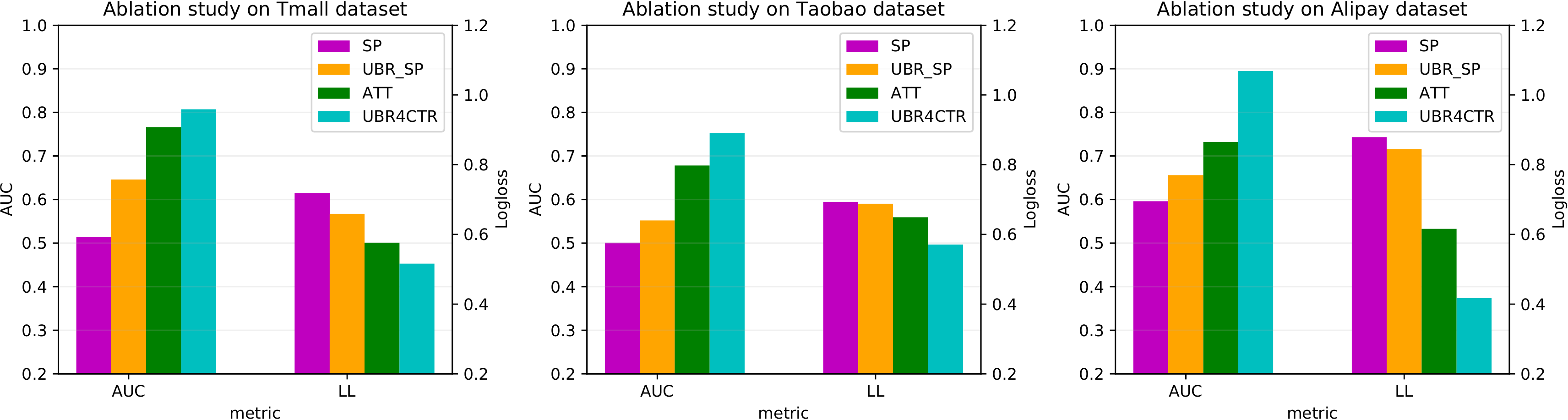}
  \caption{Ablation study on the influence of retrieval module. \emph{Note:} AUC, the higher, the better; log-loss, the lower, the better}
  \label{fig:ab_study}
\end{figure*}

To illustrate the importance of the retrieval module of our framework, we plot the performance comparisons between the sum pooling model and attention network with\&without user behaviors retrieval. 
Sum pooling model just uses a very simple sum pooling operation on user behaviors which means the $\alpha_i=1$ in Eq.~\ref{eq:weight-sum}.
The results are shown in Figure~\ref{fig:ab_study}. From the figure, we find that the sum pooling model without retrieving (SP) performs very poorly. Once equipped with the retrieval module, the performance of it increases significantly (UBR\_SP). The same phenomenon applies for attention network which performance is largely improved when equipped with behavior retrieval module (ATT vs. UBR4CTR).

\section{Deployment Feasibility}\label{sec:deploy}
The deployment of UBR4CTR has been made on the engineering schedule of a daily item recommendation platform under a mainstream bank company. In this section, we mainly discuss the feasibility of industrial deployment of UBR4CTR framework.
First, it is not difficult to switch the current model pipeline to UBR4CTR because the main change brought from UBR4CTR is the how the historical user behaviors is obtained.
To update the model pipeline to UBR4CTR, a search engine of historical user behaviors needs to be built,
while the whole CTR prediction model pipeline remain almost the same but adding an additional retrieval module. As Figure~\ref{fig:framework} shows, the prediction module is not different from that of the traditional solutions.

Efficiency is another essential concerns in industrial applications. We analyze the time complexity of UBR4CTR which is $O(T + K_q \cdot \frac{\mathcal{N}}{F})$ in Section~\ref{sec:time-comp}. For most of the sequential CTR models which are normally based on RNN, time complexity of them is $O(C \cdot T)$ where $T$ is the average length of user sequences and $C$ is the cost of one operation (e.g. GRU rolling). The time complexity of UBR4CTR is not totally infeasible because the term $\frac{\mathcal{N}}{F}$ is very closed to a constant as $F$ is a big number and it will slow the increase of this term.

From the perspective of system load, UBR4CTR is better because it does not require maintaining all the $T$ behaviors in memory which is a common practice for the traditional methods.

\begin{figure}[h]
    \centering
    \includegraphics[width=1.0\columnwidth]{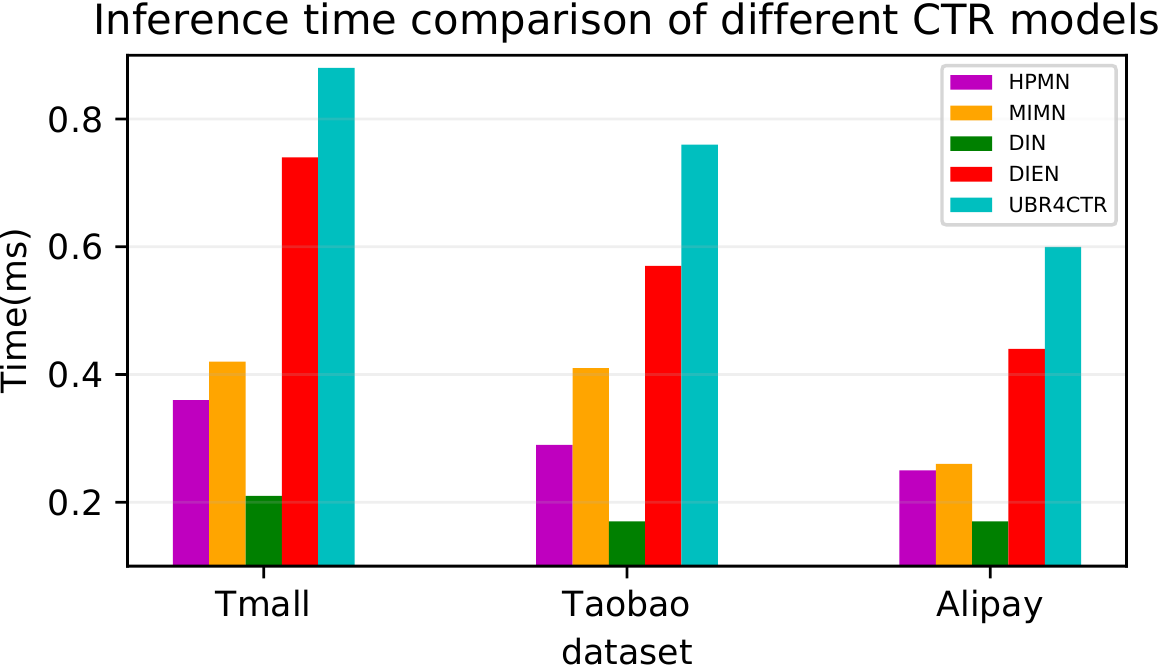}
    \caption{Inference time of different models.}
    \label{fig:time}
\end{figure}

Moreover, we compare the actual inference time between our UBR4CTR and other sequential CTR baselines in experiments. The average inference time of the models are illustrated in Figure~\ref{fig:time}. The time is calculated by dividing the overall time (only the time that contains the forward computations and behavior searching) on test dataset with the number of prediction targets.
From the figure, we could find that the absolute value of UBR4CTR's inference time on three datasets is less than 1ms which is efficient enough for online serving \cite{wang2017display}.
The inference time of UBR4CTR is the longest among all the sequential CTR models but the gap is not that large. 
Particularly, compared with DIEN which has been deployed in Alibaba online advertising platform \cite{zhou2019deep}, the average inference time of UBR4CTR is about 15\% to 30\% longer, which can be optimized via further infrastructure implementation.

\section{Related Work} \label{sec:related}
\subsection{User Response Prediction}
For user response prediction, there are two main streams of methods. One stream is about modeling the interactions of multiple categorical features. The key point of these models is to design structure that could capture the cross feature interactions. 
Factorization machine \cite{rendle2010factorization} is the pioneer model which uses matrix factorization in CTR prediction task and a lot of variants of FM \cite{juan2016field,xiao2017attentional,ta2015factorization} are proposed in recent years. Except for the early feature interaction models, deep neural networks are also used in CTR prediction. 
Wide\&Deep \cite{cheng2016wide} is the first deep learning model which effectively transform the high dimensional discrete features into dense representation and achieve good prediction performance. 
After Wide\&Deep and FM, more and more models which combine the feature interaction structure and deep neural networks are proposed. DeepFM \cite{guo2017deepfm} uses both FM and DNN to improve the CTR prediction performance. DeepCross \cite{wang2017deep} and PNN \cite{qu2016product} automatically models the cross feature interactions by outer product and inner product respectively. There are similar models such as xDeepFM \cite{lian2018xdeepfm}, FNN \cite{zhang2016deep} etc.

The other stream of models focuses more on mining temporal patterns from sequential user behaviors. DIN \cite{zhou2018deep} is an attention-based network which attributes different weights on different items that user has interacted with. DIEN \cite{zhou2019deep} utilizes two layers of GRU and attention mechanism to capture evolving user interests. HPMN \cite{ren2019lifelong} is a memory network-based method which model very long sequences for user response prediction. MIMN \cite{pi2019practice} is based on neural Turing machine \cite{graves2014neural} that models the multiple channels of user evolving interests.

\subsection{Sequential User Modeling}
Sequential user modeling is about capture user's drifting dynamics of behaviors. It is a research hotspot for recommender systems recently. Multiple types of model are proposed. The first is temporal collaborative filtering \cite{koren2009collaborative} which considers the drifting user preferences. 
The second type is based on Markov chains \cite{he2016vista,he2016fusing,rendle2010factorizing} which models the user state dynamics in an implicitly way and drive the future behaviors. 
The third category is based on deep learning. There are RNN-based models \cite{hidasi2015session,hidasi2017recurrent,wu2017recurrent,jing2017neural,liu2016context,beutel2018latent,villatel2018recurrent} that regard user behaviors as sequence of tokens, CNN-based models \cite{tang2018personalized,kang2018self} which regard the behaviors as an image and Transformer-based models \cite{kang2018self}. 
Furthermore, there are models \cite{wu2019dual} that not only utilize user-side sequence but also item-side sequence. \citet{qin2020sequential} propose dual side neighbor based CF for sequential user modeling. Memory networks are also used for sequential user modeling \cite{Ebesu:2018:CMN:3209978.3209991,wang2018neural,chen2018sequential,pi2019practice} which aim to memorize longer sequence of user behaviors.

\section{Conclusion And Future Work} \label{sec:conclusion}
In this paper, we propose the UBR4CTR framework of user response prediction. The retrieval module of our framework generates a query to search from the whole user behaviors archive to retrieve the most useful behavioral data for prediction. The retrieved data is then used by an attention-based deep network to make the final prediction. Our framework overcomes the practical problems of traditional framework which simply uses most recent behaviors and significantly improves the CTR prediction performance.
The deployment of UBR4CTR has been made on the engineering schedule of a daily item recommender system of a mainstream bank company.

For the future work of research, we will put more efforts on distributed training algorithm which will make the framework more efficient to train in a mini-batch manner. Furthermore, we will explore new effective indexing and retrieval methods for storing and searching the user behavioral archive.

\textbf{Acknowledgement}. The corresponding author Weinan Zhang thanks the support of National Natural Science Foundation of China (61702327, 61772333, 61632017).

\bibliographystyle{ACM-Reference-Format}
\balance
\bibliography{ind16}

\end{document}